\documentclass[nofootinbib,uaps,prl,reprint,showpacs,preprintnumbers,amsmath,amssymb,floatfix,a4paper,superscriptaddress]{revtex4-1}

\usepackage[T1]{fontenc}
\usepackage[utf8x]{inputenc}

\usepackage{graphicx}
\usepackage{dcolumn}
\usepackage{bm}
\usepackage{color}
\usepackage{units}
\usepackage{isotope,xspace}
\usepackage[squaren,thinspace,mediumqspace,Gray]{SIunits}

\usepackage{xcolor}
\definecolor{medium-blue}{rgb}{0,0,0.5}
\usepackage[colorlinks=true,final,unicode]{hyperref}
\hypersetup{
	pdftitle={Constraints on the Dark Matter Interpretation $n \rightarrow \chi + e^+ e^-$ of the Neutron Decay Anomaly with the PERKEO II experiment},
	pdfauthor={M. Klopf, E. Jericha, B. Märkisch, H. Saul, T. Soldner, H. Abele},
    colorlinks, linkcolor={medium-blue},
    citecolor={medium-blue}, urlcolor={medium-blue}
}

\newcolumntype{d}{D{.}{.}{-1}}

\newcommand{\perkeoII}{\textsc{Perkeo~II}\xspace}

\newcommand{\ill}{Institut Laue-Langevin, BP~156, 6,~rue Jules Horowitz, 38042 Grenoble Cedex~9, France}
\newcommand{\ati}{Atominstitut, Technische Universität Wien, Stadionallee~2, 1020 Wien, Austria}
\newcommand{\tum}{Physik-Department ENE, Technische Universität München, James-Franck-Straße 1, 85748 Garching, Germany}

\begin{document}

\preprint{2018004}

\title{Constraints on the Dark Matter Interpretation $n \rightarrow \chi + e^+ e^-$ of the Neutron Decay Anomaly  with the \perkeoII experiment}

\author{M.~Klopf}
\affiliation{\ati}



\author{E.~Jericha}
\affiliation{\ati}

\author{B.~Märkisch}
\affiliation{\tum}
\author{H.~Saul}
\affiliation{\tum}
\affiliation{\ati}

\author{T.~Soldner}
\affiliation{\ill}

\author{H.~Abele}
\affiliation{\ati}

\begin{abstract}
Discrepancies from in-beam and in-bottle type experiments measuring the neutron lifetime are on the 4$\sigma$ standard deviation level. In a recent publication Fornal and Grinstein proposed that the puzzle could be solved if the neutron would decay on the one percent level via a dark decay mode, one possible branch being $n \rightarrow \chi + e^+ e^-$. With data from the \perkeoII experiment we set limits on the branching fraction and exclude a one percent contribution for $95\,\%$ of the allowed mass range for the dark matter particle.
\end{abstract}

\date{\today}

\pacs{13.30.Ce, 12.15.Ji, 12.60.-i, 14.20.Dh}

\maketitle

Neutron decay, as the prototype for nuclear beta decay, and its
lifetime are needed to calculate most
semileptonic weak interaction processes and used as input to
search for new physics beyond the standard model of particle physics ~\cite{RevModPhys.83.1111,RevModPhys.83.1173,ABELE20081,1803.08732}. Measurements of the neutron lifetime
fall into two categories~\cite{0954-3899-16-4-001}: in the storage method neutrons
are confined in a material or magnetic bottle and after a given time
the surviving neutrons are counted. In the beta decay method, the
specific activity of an amount of neutrons (a section of a neutron
beam, a neutron pulse or stored neutrons) is measured by detecting
one of the decay products, proton or electron.
A review of neutron lifetime measurements can be found in~\cite{RevModPhys.83.1173}.
The averaged results of both categories,
879.4(6) s and 888.0(2.0) s, deviate by 8.4 s from each other,
corresponding to $4\sigma$ (all numbers from~\cite{PhysRevLett.120.202002}).

Although this lifetime discrepancy may be related to underestimated
systematics in experiments, there is a basic difference between
the two categories: the storage method measures the inclusive lifetime, independent of the decay or disappearance channel, whereas the beta decay method detects the partial lifetime into a particular decay
branch. Historically, Green and Thompson have used this argument to
derive an upper limit on the decay into a hydrogen atom which would
be missed by the beta decay method~\cite{0954-3899-16-4-001}; however, the expected
branching fraction of $4\times 10^{-6}$~\cite{Nemenov80} is too small to explain
the 8.4 s difference observed today. Greene and Geltenbort have
speculated that the discrepancy might be caused by oscillations
of neutrons into mirror neutrons~\cite{Greene16}. Recently, Fornal and
Grinstein~\cite{PhysRevLett.120.191801} have proposed different decay channels involving a
dark matter particle. These branches would have been missed by the
most precise beta decay method experiments which have detected
decay protons~\cite{Yue:2013}.

Neutron stars have been used to severely constrain these
branches~\cite{Motta2018,PhysRevLett.121.061801,PhysRevLett.121.061802} but some
models evade these constraints~\cite{1803.04961}.
Czarnecki et al. have derived a very general bound of $<0.27$\,\%{}
($95\,\%$ C.L.) on exotic decay branches of the neutron where they use
their favored values of the neutron lifetime $\tau_n$ from the
storage method and the axial coupling $g_A$ from recent beta
asymmetry measurements and assume that $V_{ud}$ from superallowed
beta decays and CKM unitarity are negligibly affected by exotic
new physics. This means that not more than 2.4 s
(with $95\,\%$ C.L.) of the lifetime discrepancy might be explained by
a dark decay. This constraint could be overcome by a smaller axial vector coupling constant $g_A$
in the range $1.268 < g_A < 1.272$~\cite{PhysRevLett.120.202002}, in
contradiction to the most accurate measurements \cite{PhysRevLett.110.172502,PhysRevC.87.032501,Markisch:2018a}. An extended anaysis can be found in~\cite{Dubbers:2019}.
We note that the interpretation of the neutron decay anomaly is relevant
for tests of the unitarity in the first row of the CKM matrix~\cite{Abele04}
with neutron decay. Recent reanalysis of the universal radiative correction
$\Delta^V_R$ to neutron and superallowed nuclear beta decay raises
tension with the CKM unitarity constraint
based on superallowed $0^+ \rightarrow 0^+$ beta decays and Kaon decays~\cite{Seng:2018a}.

Experimental constraints on the dark matter interpretation of the neutron decay anomaly have been
set on two decay branches. A recent experiment at Los Alamos National Lab~\cite{PhysRevLett.121.022505}
excludes the proposed decay channel $n \rightarrow \chi + \gamma$ as sole explanation of the
lifetime discrepancy with $97\,\%$ C.L. via a direct search for a monoenergetic $\gamma$ line.
Another decay channel, $n\rightarrow \chi + e^+ e^-$, has been searched for by the
UCNA collaboration~\cite{PhysRevC.97.052501}.
For this decay channel, the sum of the kinetic energies of the positron and electron
$E_{e^+e^-} = E_{e^+}+E_{e^-}$ is restricted to the range of $0-644\,\mathrm{keV}$,
corresponding to a dark matter mass range of between $937.900\,\mathrm{MeV}$ and
$938.543\,\mathrm{MeV}$.
The UCNA collaboration sets limits on this branching fraction of $<10^{-4}$ ($90\,\%$ C.L.) in the energy range $100\,\mathrm{keV} < E_{e^{+} e^{−}} < 644\,\mathrm{keV}$ which excludes this channel as only explanation for the lifetime discrepancy at the $5\sigma$ level \cite{PhysRevC.97.052501}.

With this letter, we set limits on the same decay channel $n\rightarrow \chi + e^+ e^-$ from data taken by the \perkeoII instrument, which was installed at the PF1B cold neutron beam position~\cite{Haese02,Abele06} at the Institut Laue-Langevin (ILL).
A drawing and a more detailed description of the \perkeoII spectrometer together with measurements of
beta decay correlation coefficients can be found
in~\cite{PhysRevLett.110.172502,PhysRevLett.100.151801,PhysRevLett.99.191803,REICH2000535}.
For the investigation of a dark decay of the neutron into an $e^{+}e^-$ -pair, we re-analyze the data that
was used to extract the beta asymmetry parameter $A$~\cite{PhysRevLett.110.172502}. In that
set-up the spectrometer is configured for electron detection only.
The electrons are transported from the decay volume towards either of the two detectors by a magnetic field of approximately
$1\,\mathrm{T}$. Details on the adiabatic transport of charged particles in magnetic fields can be found e.g.\ in~\cite{0812.0347}. For the electron detection we used two plastic scintillators each read out by four fine mesh photomultiplier tubes (PMTs). The integrated pulse size of this detection system is largely proportional to
the incident kinetic energy of a single electron or positron. During the
measurements, the detector response function was determined and the detector stability checked regularly using four monoenergetic conversion
electron sources.
The detectors showed a nonlinearity at low energy, which was
modelled for this search for a dark neutron decay mode using a quenching model developed by
Birks~\cite{birks}. 
This extends the analysis of Mund et al.~\cite{PhysRevLett.110.172502}.
The stopping power of electrons inside the scintillator material is calculated from ESTAR data~\cite{BERGER2005}. The detector calibration, including the Birks non-linearity parameter, was obtained  by a fit to the electron spectra of each detector.
The uncertainties in the non-linearity relations of both detectors are taken into account for the analysis. The energy resolution does not play an important role for the present analysis, because a variation of 50\% affects the limit on the dark matter ${e^{+} e^{−}}$ branching ratio only on the 10$^{-3}$ level.

About $8\,\mathrm{\%}$ of the electrons impinging on one
of the detectors are scattered back from the detector
and deposit only part of their energy in it. However, in the \perkeoII spectrometer such electrons
will be guided along a magnetic field line to the other detector and will, a
few nanoseconds later, deposit their energy there. For
about half of the electrons, which are backscattered near the
glancing angle, the fringe field of the magnet acts as
a magnetic mirror and projects the electron back onto
the same detector. So all electrons are confined by the
magnetic field between the two detectors and can lose
energy only to them. If, for each event, the total
signal amplitudes from both detectors are added
up, then the pure line spectrum is recovered. Details on the electron backscatter suppression can be found in~\cite{ABELE199326}.

\begin{figure}[htb]
\centering
\includegraphics[width=1.0\linewidth]{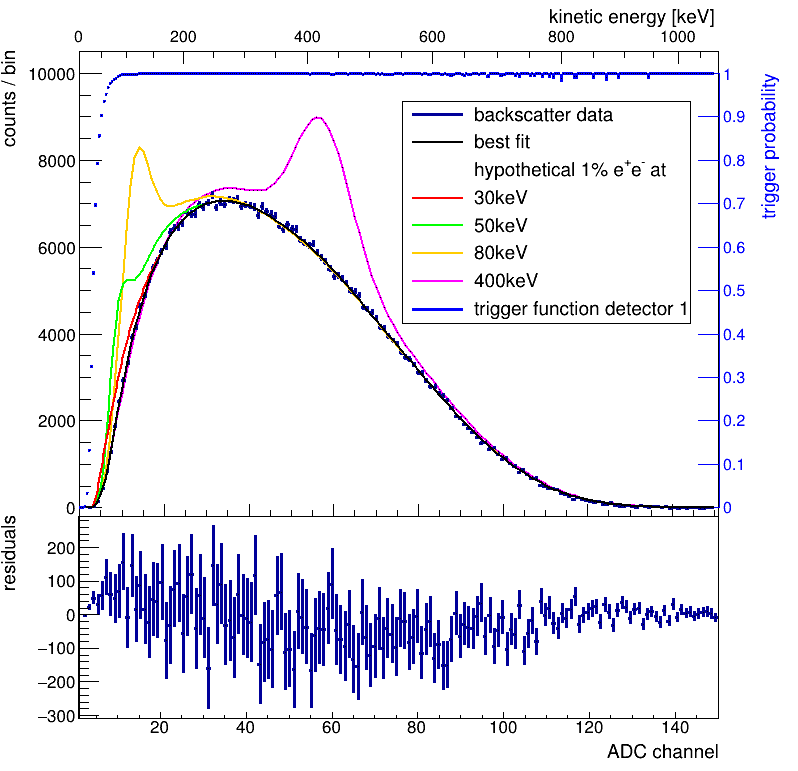}
\caption{Reanalysis of events of electrons backscattered from the \perkeoII detector system in a measurement of the
beta correlation coefficient $A$ and a search for an additional hypothetical dark matter $e^+ e^-$ signal.
Shown is a fit with residuals to the summed coincidence spectrum together with a hypothetical $1\,\%$ $e^+ e^-$ branch at $30\,\mathrm{keV}$,
$50\,\mathrm{keV}$, $80\,\mathrm{keV}$, or $400\,\mathrm{keV}$.  The backscatter signal makes up
approximately $4\,\%$ of the total beta decay events, and is the sum of events registered in both detectors. We show the trigger probability for detector 1, too. The error bars show the statistical errors.	}
 \label{fig:spectra}
\end{figure}

\begin{table}[t]
\begin{tabular}{|c|c|c|c|}
  \hline
  ADC channel & Statistical & Calibration  & Backscatter  \\
    & Error & Error & Model Error \\
  5 & 3.1 $\times 10^{-4}$ & 4.09 $\times 10^{-4}$ &8.92 $\times 10^{-4}$  \\
  10 & 1.50 $\times 10^{-4}$ & 1.40 $\times 10^{-4}$& 5.98 $\times 10^{-4}$ \\
  50& 9.28 $\times 10^{-5}$ & 9.11 $\times 10^{-5}$ &9.28 $\times 10^{-5}$ \\
  100& 4.12 $\times 10^{-5} $& 1.37 $\times 10^{-4}$& 3.47 $\times 10^{-5}$ \\
  \hline

\end{tabular}
\caption{One sigma standard deviation error budget for the fit for a  hypothetical $e^+ e^-$ dark matter
branch shown in Fig.~2 at selected channels 5, 10, 50, 100. The backscatter model error includes the
GEANT4 uncertainty in predicting the backscattering coefficient of electrons as a function of energy.}
\end{table}

The search for the proposed dark matter signal proceeded in the following way:
Most of the conventional beta decay events are rejected by requiring that both detectors
have triggered. For the remaining events, the spectrum of the total energy deposition is obtained by summing up the signals of both detectors. It is composed of conventional beta decay events with
electron backscattering, background events that trigger both detectors, and of
hypothetical $e^+e^-$ events. Background events contribute with $2\,\%$ to the
spectrum and were measured regularly with the neutron beam closed and
subtracted from the data. The $e^+e^-$ pairs are monoenergetic and would create
a characteristic peak on the backscattering spectrum, in the range from
$0\,\mathrm{keV}$ and $644\,\mathrm{keV}$ depending on the mass of the
hypothetical dark matter particle. Note that the selection cut excludes undetected
backscattering events (see~\cite{SCHUMANN2008,Roick18}) and $e^+e^-$ pairs at low energy
or going to the same detector.
Positron annihilation gamma effects are small because of the low sensitivity of the thin (5\,mm) plastic scintillators to
511\,keV gammas and are taken into account in the analysis~\cite{Klopf19}.
Tests with a $^{22}$Na positron source were performed and the effects on the expected $e^+e^-$ signal were simulated with GEANT4~\cite{AGOSTINELLI2003250}.

The expected backscatter spectrum from conventional beta decay, which
is the remaining background in the search for the hypothetical $e^+e^-$ peak, is
determined by simulations: Decay electrons are created with the angular
and energy distribution from conventional neutron decay. For each electron,
the impact angle on the detector is determined from the ratio of the
magnetic field in the \perkeoII decay volume and at the detector.

Backscattering splits the kinetic energy of an electron in two parts deposited in the two detectors. It is simulated using GEANT4~\cite{AGOSTINELLI2003250}
  with the single Coulomb scattering model, which is appropriate for low energy backscattering as this option
  reproduces experimental data above a few tens of keV. In the few keV energy range some measurements find a different backscattering fraction~\cite{1502.01507}. We take this deviation as $1\sigma$ standard deviation error on our backscatter model.
  The magnetic mirror effect for backscattered electrons is taken into account in the simulations. For the energy splitting of $e^+e^-$-events between the both detectors theoretical predictions from~\cite{1806.10107} are used. The resulting signals are obtained accounting for quenching in the scintillators, the statistical distribution of the  photo-electron conversion of the
  PMTs (which is dominating the energy  resolution), and
  additional broadening due to the noise of the charge to
  digital conversion.\\
The spectrum of the sum
  of the simulated signals still needs to be
  corrected for the trigger efficiencies of the two detectors.
  The experimental trigger efficiencies of both detectors are measured, for a signal in ADC channel $C$ of detector 1 e.g. it is
   \begin{equation}\label{trigger}
   T_{{\rm exp},1}(C) = \frac{N_1(C, 1 \& 2)}{N_1(C, 2)},
   \end{equation}
where $N_1(C, 1 \& 2)$ is the number of events in detector 1 where both detectors have triggered and $N_1(C, 2)$ the number of events in detector 1 where detector 2 has triggered. For the hardware trigger condition of detector $i$ at least two out of the four photomultipliers must have triggered.

The correction for the trigger efficiencies is obtained by applying, event by event, the trigger efficiency functions $T_{\rm{exp} ,\mathit{i}}$ to the simulated signals in the two detectors.

\begin{figure}[htb]
\centering
\includegraphics[width=1.0\linewidth]{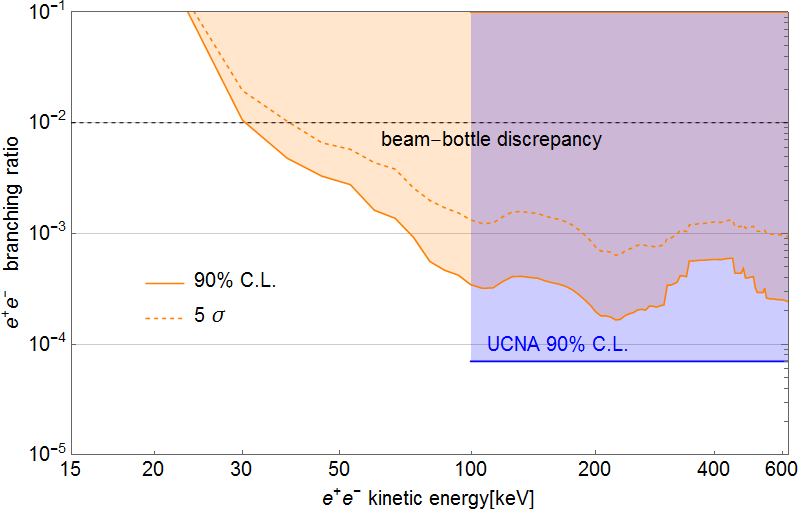}
\caption{Exclusion plot for a hypothetical $e^+ e^-$ dark matter branch in
neutron beta decay for $90\,\%$ C.L. and 5 $\sigma$ exclusion limits from a $\chi^2$ analysis. The spectrum shown in Fig.~\ref{fig:spectra} is scanned by shifting the energy of a potential $e^+ e^-$ peak in steps of one ADC channel, and performing a fit at each position. Free fitting parameter is the $e^+ e^-$ amplitude. On the $90\,\%$ confidence limit a $1\,\%$ contribution is excluded from $32\,\mathrm{keV}$ to $644\,\mathrm{keV}$, which is the maximum energy according to~\cite{PhysRevLett.120.191801}. The contributions of statistical and systematic errors are shown in the table for selected ADC channels. For comparison we show approximate results of UCNA, extracted from \cite{PhysRevC.97.052501} }
 \label{fig:Exclusion}
\end{figure}

In Fig.~\ref{fig:spectra} we show the experimental spectrum after background subtraction
together with a fit using the results of the  GEANT4 simulations.
For illustration of the signature of the hypothetical $e^+e^-$ peak also shown is the expected shape of the
spectrum for a $1\,\%$ branching to $\chi e^+e^-$ for
$e^+ e^-$ total kinetic energies of $30\,\mathrm{keV}$, $50\,\mathrm{keV}$, $80\,\mathrm{keV}$, and $400\,\mathrm{keV}$.
We scan the spectrum by shifting a
hypothetical peak in steps of one channel of the analog to digital converter (ADC), which corresponds
to approximately $6\,\mathrm{keV}$, and performing a fit at each position. The height of the $e^+ e^-$ peak is the single free parameters of the fit.
The phase space of $e^+ e^-$ pairs in the proposed dark decay has been computed in ~\cite{1806.10107}. Under the assumption of a parity conserving dark decay, the probability that the electron and positron are emitted towards opposite detectors varies  between $47.8\,\%$ and $50\,\%$, depending on the mass of the dark matter particle. As only these events can pass our selection cut, we assume the most conservative case for the $e^+ e^-$ emission of $47.8\,\%$ in the exclusion analysis. \\

In Fig.~\ref{fig:Exclusion} we show an exclusion plot for a hypothetical $\chi e^{+} e^{-}$ branching fraction at
$90\,\%$ C.L., which corresponds to a one sided 1.3 $\sigma$ cut above the best fit value for the dark matter
branching ratio.
In regions, where the fit to the amplitude of the dark matter signal has a negative outcome, we renormalize the tail probabilities in the positive range and take a $90\,\%$ cut. We also tested the significance of observing a local
excess of events, i.e.\ a dark matter signal somewhere
in a possible mass range if we take into account the probability of
observing such an excess anywhere in the range. We quantify this ``look-elsewhere effect'' by obtaining $10^7$ Monte-Carlo
simulations of backscatter only data, and for
each channel the largest fluctuation that resembles a signal.
Details on the ``look-elsewhere effect'' can be found e.g.\ in~\cite{Gross2010}.
The proposed $1\,\%$ contribution to neutron beta decay~\cite{PhysRevLett.120.191801}
can be excluded with $90\,\mathrm{\%}$ C.L. between  $32\,\mathrm{keV}$ and $664\,\mathrm{keV}$ and better than 5 sigma standard deviation for energies between $37.5\,\mathrm{keV}$ and $664\,\mathrm{keV}$. This
corresponds to $95\,\%$ respectively $94\,\%$ of the allowed mass range for the dark matter particle. In
general we can derive limits for higher energies, which are, however, excluded by~\cite{PhysRevLett.120.191801}.

\begin{acknowledgments}
\begin{center}
ACKNOWLEDGEMENTS
\end{center}

We thank D. Schirra, C. Roick (Technische Universität München) and A. Ivanov (Technische Universität Wien) for useful discussions and F. Müller (Technische Universität Wien) for help with MC simulations. This work was supported by the Austrian Science Fund (FWF) contract nos. P~26630-N20, P~26781-N20 and I~689-N16,
and the DFG Priority Programme SPP 1491, contract nos. AB~128/5-2, MA~4944/1-2.
The computational results presented have been achieved in part using the Vienna Scientific Cluster.
\end{acknowledgments}

%

\end{document}